\documentclass[preprint2,twoside]{hwo}

\usepackage{graphicx}
\usepackage{longtable}
\usepackage{multirow}

\newcommand{\mstar}{$M_*$}

\newcommand{\mstareq}{M_*}

\newcommand{\msuneq}{M_{\odot}}
\newcommand{\lya}{Ly$\alpha$}

\input{hwo.h}

\begin{document}

\title{\textbf{\LARGE Mapping Galactic Winds and Small-scale Structure in the Circumgalactic Medium with Habitable Worlds Observatory}}
\author {\textbf{\large Joseph N. Burchett$^{1}$, Deborah M. Lokhorst$^2$, Yakov Faerman$^3$, Kevin France$^4$, Kate H. R. Rubin$^5$, David S. N. Rupke$^6$, and Sanchayeeta Borthakur$^7$}}
\affil{$^1$\small\it New Mexico State University, Department of Astronomy, 1320 Frenger Mall, Las Cruces, NM 88003-8001, USA}
\affil{$^2$\small\it NRC Herzberg Astronomy \& Astrophysics Research Centre, 5071 West Saanich Road, Victoria, BC V9E2E7, Canada}
\affil{$^3$\small\it University of Washington, Department of Astronomy, Seattle, WA 98195, USA}
\affil{$^4$\small\it Laboratory for Atmospheric and Space Physics, University of Colorado Boulder, Boulder, CO 80309, USA}
\affil{$^5$\small\it Department of Astronomy, San Diego State University, San Diego, CA 92182 USA}
\affil{$^6$\small\it Department of Physics, Rhodes College, 2000 N. Parkway, Memphis, TN 38104, USA}
\affil{$^7$\small\it School of Earth \& Space Exploration, 
Arizona State University, 781 Terrace Mall, Tempe, AZ 85287, USA}

% Please add the names of endorsers in the format "Joseph Jensen (Utah Valley University), " separated by commas.
\author{\footnotesize{\bf Endorsed by:}
Michelle	Berg	(Texas Christian University),
Greg	Bryan	(Columbia University),
Alison Coil	(University of California San Diego), 
Romeel	Dav\'e	(University of Edinburgh),
Michael	Davis	(Southwest Research Institute),
Rajeshwari	Dutta	(Inter-University Centre for Astronomy and Astrophysics),
Sophia	Flury	(University of Edinburgh),
Andy	Fox	(STScI),
Erika	Hamden	(University of Arizona),
Farhanul	Hasan	(Space Telescope Science Institute),
J. Christopher	Howk	(University of Notre Dame),
Oppor	Joshua	(The University of Texas at Austin),
Glenn	Kacprzak	(Swinburne University of Technology),
Brad	Koplitz	(Arizona State University),
Ralph	Kraft	(Smithsonian Astrophysical Observatory),
Dhanesh	Krishnarao	(Colorado College),
Varsha	Kulkarni	(University of South Carolina),
Alex	Lazarian	(University of Wisconsin-Madison),
Eunjeong Lee	(EisKosmos (CROASAEN, Inc.)),
Nicolas	Lehner	(University of Notre Dame),
Alexander	Menchaca	(New Mexico State University),
Drew	Miles	(California Institute of Technology),
Dylan	Nelson	(Heidelberg University),
Nikole	Nielsen	(University of Oklahoma),
Marc	Rafelski	(STScI),
Kate	Rowlands	(STScI),
Grace	Telford	(Princeton University),
Sarah	Tuttle	(University of Washington, Seattle),
Sylvain	Veilleux	(University of Maryland, College Park),
Siyao	Xu	(University of Florida),
Yong	Zheng	(Rensselaer Polytechnic Institute)
}

% This section is for ADS Processing.  There must be one line per author. Leave them commented out for the present. They will be included later.
%\paperauthor{Sample~Author1}{Author1Email@email.edu}{ORCID_Or_Blank}{Author1 Institution}{Author1 Department}{City}{State/Province}{Postal Code}{Country}
%\paperauthor{Sample~Author2}{Author2Email@email.edu}{ORCID_Or_Blank}{Author2 Institution}{Author2 Department}{City}{State/Province}{Postal Code}{Country}
%\paperauthor{Sample~Author3}{Author3Email@email.edu}{ORCID_Or_Blank}{Author3 Institution}{Author3 Department}{City}{State/Province}{Postal Code}{Country}

% Please provide entries for the Author index; leave them commented out for now.
%\aindex{Pacucci, F.}

\begin{abstract}
  We present a science case for the Habitable Worlds Observatory (HWO) to map the circumgalactic medium (CGM) in emission by targeting ultraviolet emission lines, which trace the $10^4 - 10^6$ K gas engaged in the feedback and accretion mechanisms driving galaxy evolution. While the CGM-galaxy connection is clearly evident through absorption line experiments and limited work done in optical and radio emission from the ground, the nature of this connection is poorly understood with regard to how these cosmic ecosystems exchange matter and energy.  We outline a two-pronged experiment with Habitable Worlds Observatory (HWO) utilizing both multi-object spectroscopy to map kpc-scale CGM structures such as galactic superwinds and integral field spectroscopy to unveil the sub-kpc-scale processes such as thermal instabilities, which are theorized to govern the cool gas reservoirs long detected in pencil-beam absorption surveys.\textit{This article is an adaptation of a science case document developed for HWO's CGM/IGM Steering Committee.}
\\
\\ 
\end{abstract}

\vspace{2cm}

\section{Science Goal}

\noindent\textbf{Map the interplay of gas between the CGM and galaxies to determine how feedback and accretion control the growth of galaxies
}

\begin{figure*}
\begin{center}
\begin{tabular}{c}
\includegraphics{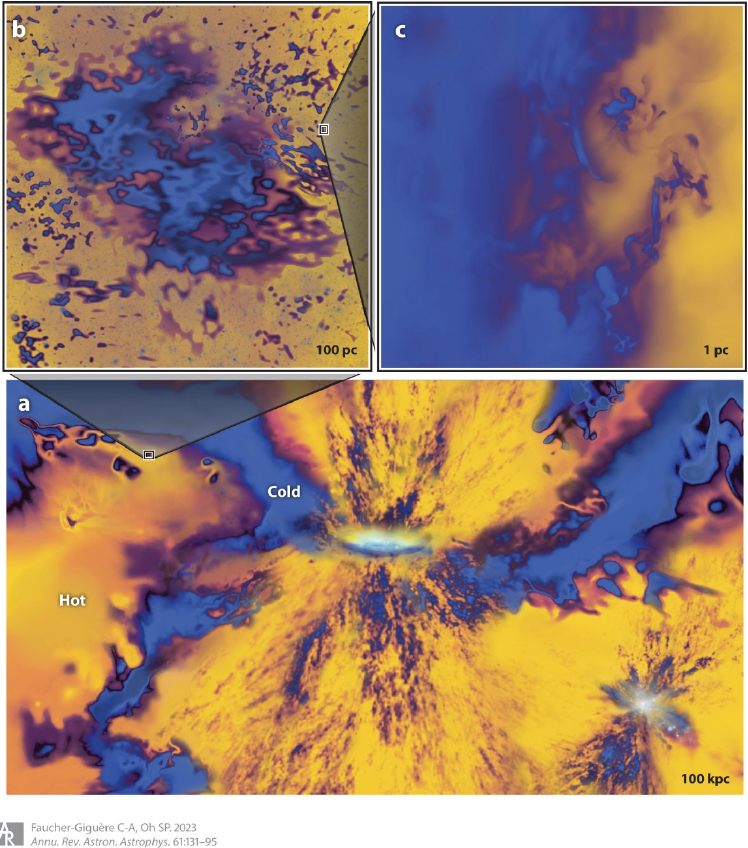}
\end{tabular}
\end{center}
\caption 
{ \label{fig:cgmScales}
Adapted from \citep{Faucher-GiguereOh:2023_cgmReview}, an artistic representation of the wide spatial dynamic range of important processes in the CGM. The bottom panel depicts large-scale galactic winds and accretion flows, structured on scales of 10s of kpc.  The upper left panel highlights an individual cold ($10^4$ K) cloud complex, $\sim$100 pc across, embedded in the ambient hotter ($10^6$ K) CGM.  A mixing boundary layer is shown to the upper right, where fractal structure continues below pc scales. In all panels, the color scale encodes the temperature, with $10^4$ K gas in blue and $10^6$ K gas in yellow.} This Science Case adopts a two-pronged approach to map, in emission, CGM structures commensurate with the bottom and upper left panels, revealing both the large scale flows driven by stellar and AGN feedback as well as cold cloud formation from the ambient hot halo.  
\end{figure*}

For half a century now, the galaxy formation and evolution community has recognized the critical role of gas accretion to sustain star formation \citep[e.g.,][]{Larson:1972gf,White:1978aa}. Gas is thought to transition into galaxies through accretion processes and, in turn, get expelled from galaxies through feedback processes, an interplay that regulates galactic gas content. What is now broadly called ‘feedback’ includes the ejection of both matter and energy from stellar winds, radiation, and supernovae as well as active galactic nuclei (AGN), and both theory and observation have implicated feedback as central to regulating, perhaps even halting (or quenching), star formation.  The discovery of a ubiquitous circumgalactic medium (CGM; Figure \ref{fig:cgmScales}) from quasar absorption line studies supports this paradigm: 1) the dark matter halos in which galaxies reside contain a substantial baryonic component well beyond any stellar extent of a galaxy (a potential source of fuel for star formation) \citep{Boksenberg:1978fk}; 2) the gaseous halos permeating their dark matter counterparts are correlated with their host galaxy star formation properties \citep[connecting the CGM to central galaxy activity;][]{Tchernyshyov:2023_o6quenching,Tumlinson:2011kx}; 3) galaxy environment impacts the gas reservoirs of galaxies living in overdense environments such as groups and clusters \citep[where quenched galaxies reside in great numbers;][]{Burchett:2018aa}; and 4) more than half of the metals in the universe reside outside galaxies, which implicates ejection from sites of star formation (galaxies) into the CGM \citep{Peeples:2014rf}.  

The CGM and intergalactic medium (IGM) have been primarily investigated through absorption line measurements because these media are too diffuse and faintly emitting to be directly measured through emission with existing observational facilities except in rare cases \citep{Burchett:2021_mg2emission,Nielsen:2024_cgmEmission,Hayes:2016_o6emission}. Absorption line measurements provide only limited spatial information, as this method relies on background sources, e.g., quasars, that are limited in number to only one sightline per galaxy and a few sightlines around only the rarest of galaxies. Emission line measurements stand to transform our view of galaxies and their ecosystems by enabling the first-ever picture of the morphology, kinematics, and phase structure of the CGM. Furthermore, emission observations overlapping with galaxies that have absorption line measurements can provide critical context for the absorption line measurements that have been hard won over the past 50 years, enabling unprecedented physical inference.  For example, the combination of emission and absorption line measurements has the unique power to constrain the gas volume density -- a quantity that cannot be unambiguously constrained by either tracer alone.  Constraining the gas density is critical to estimating the total gas mass and rates of matter and energy flows, which are in turn fundamental to understanding the baryon cycle.

Lastly, we conclude this introduction by reiterating the U.S. astrophysics community’s acknowledgment of the importance of the CGM to galaxy formation and evolution.  As stated in the Astro2020 Decadal Survey \citep{decadalSurvey:2021}:

\begin{quote}
    "The upcoming observations with JWST, the Rubin Observatory, and Roman will be profound but will not on their own be able to address the central problem of understanding how galaxies grow. Probing the heart of the galactic feedback process requires detecting and measuring the tenuous gases at the boundaries of galaxies and their intergalactic surroundings, the circumgalactic medium (CGM), where the accretion and recycling of gas and metals from feedback processes take place. This goal motivates the priority area within Cosmic Ecosystems."
\end{quote}

In particular, this Science Case is relevant to the following Key Science Question and Discovery Areas of the Astro2020 Decadal Survey Report:\\
\\
    \noindent\textbf{D-Q2.} \textbf{What Seeds Supermassive Black Holes and How Do They Grow?}
    \begin{itemize}
        \item \textbf{D-Q2a.} The Acquisition of the Gas Necessary to Fuel Star Formation
        \item \textbf{D-Q2b.} The Production, Distribution, and Cycling of Metals
        \item \textbf{D-Q2d.} The Physical Conditions of the Circumgalactic Medium
    \end{itemize}

\section{Science Objective}
\label{sect:sciObj} 

\noindent\textbf{Key Objective: What is the morphology and spatial distribution of the CGM in connection with feedback from galaxies?}

\begin{figure*}
\begin{center}
\begin{tabular}{c}
\includegraphics[width=\linewidth]{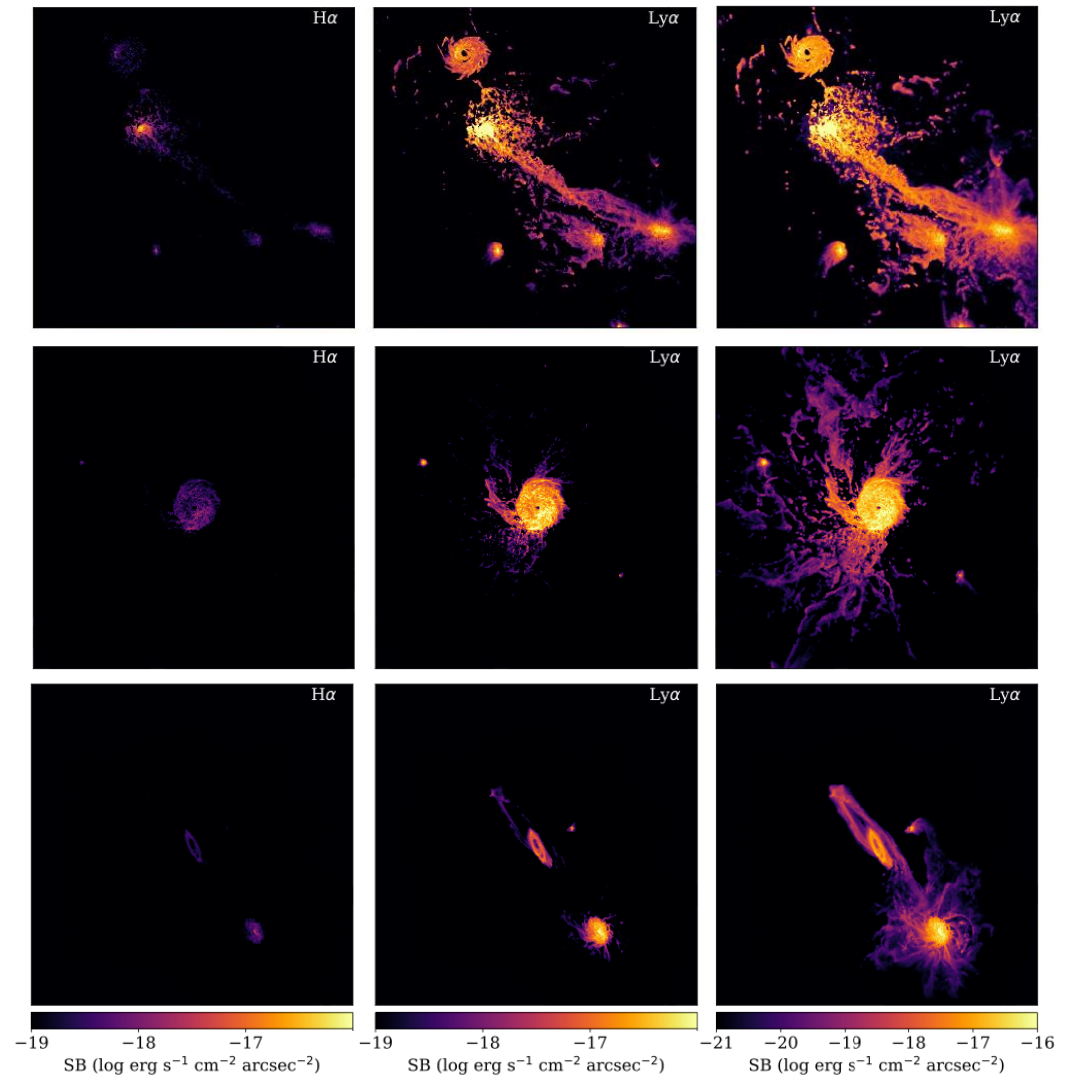}
\end{tabular}
\end{center}
\caption 
{ \label{fig:tng_halpha_lyalpha}
Simulated emission from hydrogen gas in the surroundings of three galactic systems within the TNG50 simulations. These maps are 700 kpc on a side, whereas galaxies have a typical extent in starlight of $\sim20$ kpc. At low redshifts, H$\alpha$ emission (left) would be visible from the ground, unlike \lya\ (center and right), which is emitted in the ultraviolet and must be observed from space at $z < 1.5$.  Furthermore, H$\alpha$ is highly concentrated in the central regions, while \lya\ directly traces the flow of material among galaxies. The \lya\ maps are shown with two sensitivity limits highlighting the additional structure revealed with deeper observations. As seen here, \lya\ is up to 100x brighter than H$\alpha$ emission and affords mapping the diffuse gas interplay in this cosmic ecosystem. } 
\end{figure*}

New ground-based integral field spectrographs (IFS), such as KCWI and MUSE, have come online in the past $\sim$10 years and have indeed detected the CGM in emission maps. However, some qualification is necessary here. Most of the CGM systems detected in emission are observed in H~I Ly$\alpha$ \citep[rest $\lambda$ 1215.67 \AA;][]{Cantalupo:2014aa,Cai:2019_lyaZ2}; and Mg~II  \citep[rest $\lambda\lambda$ 2796.35, 2803.53 \AA;][]{Burchett:2021_mg2emission,Zabl:2021_mg2emission}; which both trace $10^4$ K gas and shift into the visible wavelength regime only at $z=2$ and $z=0.3$, respectively (assuming enough throughput at the blue end of the bandpass, which is not the case for VLT/MUSE).  At these redshifts, relatively little can be known about the host galaxies themselves compared to more local systems at $z\sim0$, where current or soon-to-come facilities will map the molecular ($\sim$10~K; ALMA \cite{WoottenThompson:2009_ALMA}, ngVLA \cite{Murphy:2018_ngVLA_science}), atomic (100-1000~K; SKA \cite{Dewdney:2009_SKA}, ngVLA \cite{Murphy:2018_ngVLA_science}),  and hot ($>10^6$~K; Chandra \cite{Weisskopf:2002_ChandraOverview}, AXIS \cite{Reynolds:2023_AXIS}) gas as well as stellar populations and dust content on sub-kpc scales at radio, optical, and X-ray wavelengths.  Alternatively, much progress has been made to characterize faint, extended CGM emission around ensembles of galaxies through stacking \citep{Guo:2023_mg2emission,Dutta:2024_metalEmissionZ1,Dutta:2024_metalEmissionZlt1}, but these studies compromise galaxy-to-galaxy variations within stacked subsamples and wash out detailed kinematic and morphological substructure. Recent detections of extended [O II], [O III], and hydrogen Balmer series line emission are encouraging for mapping rest-frame optical line emission in the CGM \citep{Nielsen:2024_cgmEmission,Johnson:2022_emissionFilamentary,lokh22a}, but the host galaxies in these observations are rather atypical (e.g., are starbursts or quasars). Mapping CGM emission from a wide representation of galaxies requires sensitivity to ultraviolet (UV) line emission, which is $\sim10 - 100 \times$ brighter than visible wavelength line emission (see Figure \ref{fig:tng_halpha_lyalpha} for an example comparison between H$\alpha$ and Ly$\alpha$).  Figure~\ref{fig:cgmScales} highlights the dynamic range in physical scale over which salient processes occur in the CGM. X-ray observatories launching in the next 20 years (Athena \cite{Barret:2020_athenaObservatory}, AXIS \cite{Reynolds:2023_AXIS}) may well detect the hottest CGM material at $T\gtrsim10^6$ K via, e.g., the O VII and O VIII lines at 0.57 keV and 0.65 keV, respectively.  However, the complex gas kinematics will be greatly under resolved (e.g., the Athena spectrometer has a planned velocity resolution of $>2000$ km/s), and the UV is uniquely sensitive to diffuse $10^4$ K $\leq T < 10^6$ K gas \cite{Spitzer:1956_galacticCoronae}.   Completing the multi-wavelength, high-resolution view of galaxies obtained from the ground-based efforts mentioned above will require mapping the key UV CGM gas diagnostics at low redshifts, which will in turn require a spectral imaging instrument on a space telescope (Habitable Worlds Observatory) in order to detect UV emission lines at wavelengths that are blocked by Earth’s atmosphere.

Habitable Worlds Observatory will enable mapping the CGM of individual low redshift galaxies in both hydrogen and metal UV emission lines to provide a multiphase picture of the CGM.  Detailed studies of a few individual galaxies in emission can provide insight that millions of absorption data points compiled from the CGM of millions of galaxies cannot: a spatial map of the interaction between the CGM and the galaxy, linking feedback and accretion together with respect to specific galactic properties (star formation rate, or SFR; mass; stellar age; etc.).
This science case requires an observational campaign targeting low redshift galaxies to map out the circumgalactic medium for each individual galaxy \citep[as opposed to stacking;][]{Guo:2023_mg2emission}. In addition to mapping the intensity of the line emission, the velocity of the gas will also be measured; in the following discussion, it is implied that kinematic information will accompany the spatial information, constraining, e.g,. whether the gas is inflowing or outflowing.  Furthermore, emission lines from several ionic species will be mapped across the CGM, enabling one to determine temperatures, densities, metallicities, ionization state, and source of excitation  with the help of ionization models. The $<10^6$ K gas phase structure (via temperature and density) and metallicities can be inferred by mapping H I Ly$\alpha$ 1216 \AA; O VI 1032,1038 \AA; C III 977 \AA; C IV 1548,1551 \AA; Si II 1260 \AA~(plus other lines covered); Si III 1206 \AA; O I 1302 \AA; and Mg II 2796,2803 \AA\ emission.

As mentioned above, CGM emission is now being mapped in optical lines from the ground by 10-m class telescopes; however, we should highlight the radically different density regimes afforded by UV lines versus the optical.  Figure \ref{fig:tng_halpha_lyalpha} shows H$\alpha$ (optical, $\lambda_{\rm rest} = 6564$ \AA) and Ly$\alpha$ (UV, $\lambda_{\rm rest} = 1216$ \AA) surface brightness maps from the TNG50 \citep{Nelson:2019_tngDataRelease,Pillepich:2019_tng50, Nelson:2019_outflows} cosmological hydrodynamical simulation of the CGM of Milky Way-like galaxies. The data in these maps were produced using the IllustrisTNG Galaxy and Halo visualization tool\footnote{https://www.tng-project.org/data/vis/} and scaled to have consistent color mapping.  The maps do not include radiative transfer effects.  Here, the Ly$\alpha$ is predicted to be $\sim100$x brighter than H$\alpha$.  A mix of gas phases are present in the simulations. The hydrogen emission tracks mainly cool gas ($10^4$ K), which can be roughly split into dense bright regions (as seen in the left panel) and diffuse, fainter gas (as seen in the middle and right panels).  Note the dramatically improved view of this cosmic ecosystem afforded by the UV Ly$\alpha$ emission over the optical H$\alpha$ emission.  While both of these regimes are important to characterize, the UV is required to complete the picture of cosmic ecosystems in action via the exchange of matter and energy in the diffuse gas.  These same simulations predict that UV metal lines (e.g., O VI) key to analyzing the thermodynamic phase structure (via ionization modeling) should be on the order of 10-100x fainter than Ly$\alpha$ (but still feasibly detected), further underscoring the power of the UV. We provide a quantitative analysis of observational parameters in Section \ref{sect:descObs}. 

While \lya\ is likely to be the most intrinsically bright UV emission line and holds vital information on the total hydrogen content, modulo ionization corrections, the array of metal ion tracers listed above provide critical information on the metallicity and phase structure of the gas.  O I traces the neutral gas, as it has a nearly equal ionization potential to H I.  Low-ionization species Mg II and Si II trace the cool, dense $10^4$ K gas. Under collisional ionization, intermediate-ionization species Si III, C III, and C IV trace warmer gas to $\gtrsim10^5$ K; and high-ionization O VI lines mark warm-hot $\gtrsim10^{5.5}$ K gas, coinciding with the peak of the radiative cooling function for metal-enriched plasma \citep{Sutherland:1993aa}.  Figure 6 of the review by \citet{Tumlinson:2017aa} provides a handy reference of the phase space traced by several UV transitions.   

\subsection{Science Objective 1: Directly mapping galactic feedback through the CGM}

\noindent\fbox{%
    \parbox{0.49\textwidth}{%
        \textbf{Synopsis:} Galactic winds are critical to galaxy formation/evolution as we know it and appear to be ubiquitous even given our limited observational capability.  We cannot currently map them for reasonable statistical samples of normal galaxies, but certain remarkable cases can guide expectations for relevant size and velocity scales.  Theoretical predictions for mapping CGM structures such as winds vary depending on the model and may be readily tested with HWO given requisite sensitivity, FOV, and velocity resolution.
    }%
} \\  \\

The impact of galactic feedback is on display front and center in our own Galaxy \citep{BlandHawthorn:2003_bipolarWindGC}.  High-energy observatories such as the Fermi Gamma Ray telescope and the eROSITA survey have imaged enormous structures (commonly known as the Fermi Bubbles or eROSITA bubbles) extending from the center of the Milky Way to several kpc off the plane of the Galaxy.  Observers have also mapped emission around other galaxies clearly showing outflows \citep{Burchett:2021_mg2emission}, with the nearby M82 \citep{BlandTully:1988_m82} being a particularly dramatic example. Indeed, galaxy formation theory requires feedback/outflows and gas recycling to reproduce the observed galaxy population global statistics, such as the stellar mass function \citep[e.g.,][]{Oppenheimer:2010lq}. Recent state-of-the-art hydrodynamical simulations predict Fermi bubble-like structures to be quite common around galaxies \citep{Pillepich:2021_x-rayBubbles}, suggesting such outflows should be directly observable.  Furthermore, outflows appear to be ubiquitous in star-forming galaxies as revealed by absorption against the galaxy stellar continua \citep{Rubin:2014fk,Veilleux:2020_coolOutflowReview}.  The capability of mapping the morphology of galactic winds would directly probe the nature of galactic flows such that is not possible from absorption studies. 

\begin{figure*}
\begin{center}
\begin{tabular}{c}

\includegraphics[width=0.97\linewidth]{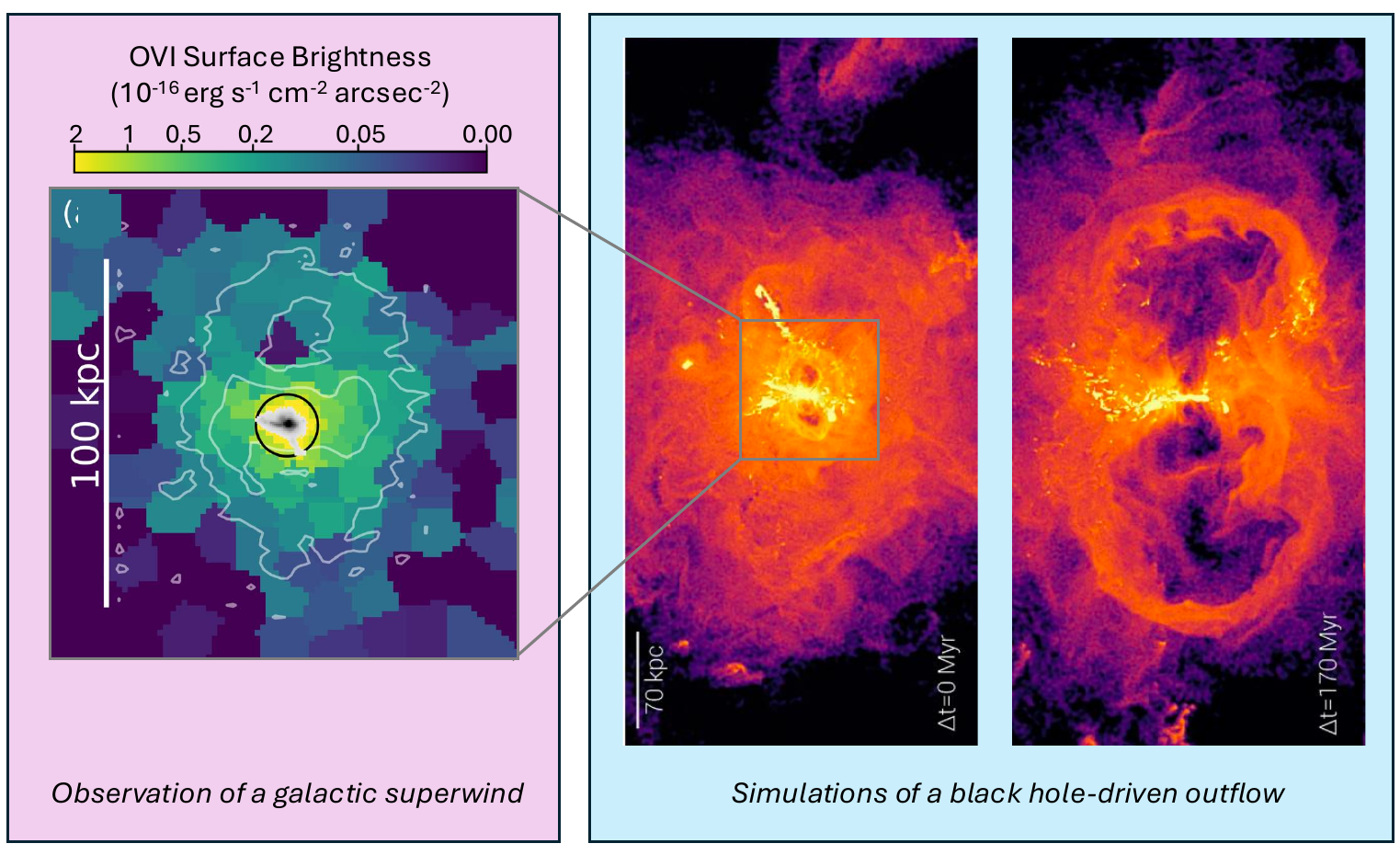}
\end{tabular}
\end{center}
\caption 
{ \label{fig:makani}
‘Makani’, a $z\sim0.5$ galactic superwind originally mapped by \citet{Rupke:2019_makani} in [O II] emission, shown in contours, and more recently in O VI emission via HST narrowband imaging by \citet[][figure adapted here]{Ha:2025_makaniOVI}}, shown by the surface brightness color scale (left panel). This bipolar structure extends over nearly 100 kpc, which sets the physical scale over which the CGM should be mapped in emission according to this science case. The right panel shows an analogous simulated galaxy from Illustris TNG50 with supermassive black hole (SMBH) driven outflows at two timesteps, demonstrating the impact of winds on the galaxies and surrounding gas with time \cite{Nelson:2019_tngDataRelease}. In these simulations, SMBH-driven feedback plays a dominant role in quenching galaxies. With the UV multi-object spectrograph, HWO could map such a structure in rest-frame UV lines tracing multiple phases of gas from $10^4-10^6$ K in 1-3 pointings depending on the target redshift and field-of-view of the instrument (see Table \ref{tab:physPars} and Section \ref{sect:descObs}).  Furthermore, HWO spectroscopy would deliver kinematics of the gas, which is missing from the HST narrowband data.  We emphasize the brightness of the detected emission, which exceeds $10^{-17}$ erg s$^{-1}$ cm$^{-2}$ arcsec$^{-1}$ over $\sim50$ kpc scales and exceeds that predicted by the simulations shown in Figure \ref{fig:sim_sn_maps}, although we note the galaxies differ in \mstar\ and SFR.  This empirical result lends further confidence to the success of HWO achieving the science objectives herein.   
\end{figure*} 

Certain remarkable cases observed with ground-based instrumentation can guide observing strategies for more prolific space-based investigations of galactic flows via emission.  The ‘Makani’ object (Figure \ref{fig:makani}; \citet{Rupke:2019_makani,Rupke:2023_ionizationDynamicsMakani})  was discovered as a galactic superwind exhibiting bipolar emission in [OII], [OIII], and other optical lines over  $\sim$100 kpc. The wind has two outflow components at 300 km/s and 1000 km/s, which map to previous starburst episodes identified in the star formation history.  \citet{Ha:2025_makaniOVI} imaged Makani in O VI line emission using a synthetic narrowband imaging technique \citep[see][]{Hayes:2016_o6emission}, revealing that the O VI surface brightness exceeded that of [O II] by factors of $\sim$2--10$\times$ in the outer regions.  Of course, imaging alone does not reveal kinematics, and spectrally mapping the FUV line emission (lines listed above), which should generally be 10-100$\times$ brighter than the optical as also suggested by Figure \ref{fig:tng_halpha_lyalpha}, would enable inferring the density, temperature, metallicity, and mass of the gas across multiple phases from T = $10^4-10^6$ K.  Indeed, the high O VI surface brightness measured by \citet{Ha:2025_makaniOVI} suggests that not only will such detections be well within reach of HWO, but HWO spectroscopy could characterize the line emission and reveal the kinematics of the gas unavailable from narrowband imaging. A KCWI study of 11 other compact starburst galaxies (plus Makani) revealed extended optical line nebulae around all objects targeted, with typical scales of approximately 40 kpc \citep{Perotta:2024_outflowingNebulae}.  An HWO UV MOS could map winds on these scales in one pointing at $z\sim0.01$ given a field-of-view of 8'x8' (see Figure \ref{fig:mos_fov_res}) with $\lesssim$ kpc-scale spatial resolution.

Such structures in the CGM may be more commonplace around more typical galaxies (not just those with exceptionally high SFR density), and surveys targeting the CGM of ‘normal’ galaxies are warranted.  \citet{Pillepich:2021_x-rayBubbles} found Fermi bubble-like structures around 2/3 of Milky Way- and Andromeda-like galaxies in the IllustrisTNG simulations.  From simulations of a Milky Way-like galaxy, \citet{Corlies:2016aa} predict metal-line emission from O VI exceeding surface brightnesses of 10$^{-19}$ erg s$^{-1}$ cm$^{-2}$ arcsec$^{-2}$, well within reach of HWO, on 100 kpc scales.  Notably, these and other simulations \citep{Oppenheimer:2016lr} underpredict the column densities observed in absorption studies \citep{Tumlinson:2011kx,Werk:2013qy}, indicating that the observed emission may even exceed their predicted surface brightnesses.  The high covering fractions of absorbers tracing cool and warm-hot gas within 100 kpc of galaxies \citep[see][and references therein]{Tumlinson:2017aa} underscore the ubiquity of enriched gas in the CGM, suggesting a high likelihood of detecting CGM emission around large samples of galaxies given sufficient sensitivity.  \citet{Piacitelli:2022_cgmEmission} used the column densities directly from absorption measurements to predict the emission from various UV lines and find that the surface brightness of O VI may exceed the predictions of \citet{Corlies:2016aa} by $\sim2$~dex.   Further empirical evidence comes from the recent O VI emission map from  \citet{Ha:2025_makaniOVI} reporting surface brightness $>10^{-17}$ erg s$^{-1}$ cm$^{-2}$ arcsec$^{-2}$ over 50 kpc scales (see Figure \ref{fig:makani}).

\subsection{Science Objective 2: Unveiling thermodynamic instabilities in the CGM} 

\noindent\fbox{%
    \parbox{0.49\textwidth}{%
        \textbf{Synopsis:} A key mechanism by which galaxies may accrete gas from their massive circumgalactic reservoirs is through condensation out of the hot halo into cold gas substructures.  This process is predicted to form much smaller structures than we can currently map in emission.  Although it possible to detect the cool gas through low-ionization UV tracers in absorption, these lack spatial information necessary to test models, which produce different characteristic scales under different CGM conditions.  HWO can achieve the spatial resolution to detect such small scale structures.
    }%
} \\  \\

HWO stands to rectify a long-standing disconnect between theory and observation.  Analytical theory predicts that the mechanisms that produce cold gas in the CGM act on scales of just a few pc -- scales that even state-of-the-art cosmological simulations cannot resolve.  While absorption-line observations are sensitive to this gas, they offer only very weak constraints on its small-scale structure, meaning that current UV instrumentation cannot be used to test these theories of cold gas formation.  For the first time, we will have the opportunity to map the relevant scales (few pc) in UV emission with HWO, particularly if equipped with a sub-arcsecond resolution integral field unit (IFU).

\citet{McCourt:2018aa} formulated a useful `baseline' of how gas forms substructure in the CGM through hydrodynamic instabilities.  The mechanism known as ‘shattering’ leads to the condensation of gas into a ‘mist’ or ‘fog’ that surrounds galaxies, where diffuse structures such as the gaseous halos of galaxies readily detected in absorption with HST on scales of 10s-100s of kpc are composed of sub-kpc scale cloudlets. 

Following \citet{McCourt:2018aa}, a key metric used to quantify the typical scale of cold gas complexes formed through radiative cooling is the cooling length, defined as the sound speed times the cooling time: $L_{\rm cool} = v_s \times t_{\rm cool}$.  The shattering scale is defined as min($L_{\rm cool}$) at some characteristic density, which may be approximated as follows:

\begin{equation}
    L_{\rm shattering} = {\rm min}(v_s \times t_{\rm cool}) \sim 0.1~{\rm pc}\left(\frac{n}{\rm cm^{-3}}\right)
    % L_{\rm shattering} = {\rm min}(v_s \times t_{\rm cool}) \sim 10~{\rm pc}\left(\frac{n}{\rm 10^{-2}~cm^{-3}}\right)
\end{equation}

 Deviations from this idealized (thermal) shattering scenario may lead to order-of-magnitude differences in the characteristic sizes of small-scale cloudlets. On one hand, hydrodynamic processes, such as turbulence and instabilities (Kelvin-Helmholtz or Rayleigh-Taylor, for example), can disrupt and fragment the clouds to smaller sizes than predicted by shattering. On the other hand,  magnetic fields can suppress such instabilities and shield the clouds from disruption \citep{Jung:2023_Bfields}, and cosmic rays can provide extra pressure \citep{Ji:2019_CRs}, keeping the gas clouds from contracting. Both processes can result in smoother cool gas morphologies \citep{Butsky:2020_cosmicRaysThermalInstab,Tsung:2023_cosmicRaysStability}.

\begin{figure*}
\begin{center}
\begin{tabular}{c}
\includegraphics[width=\linewidth]{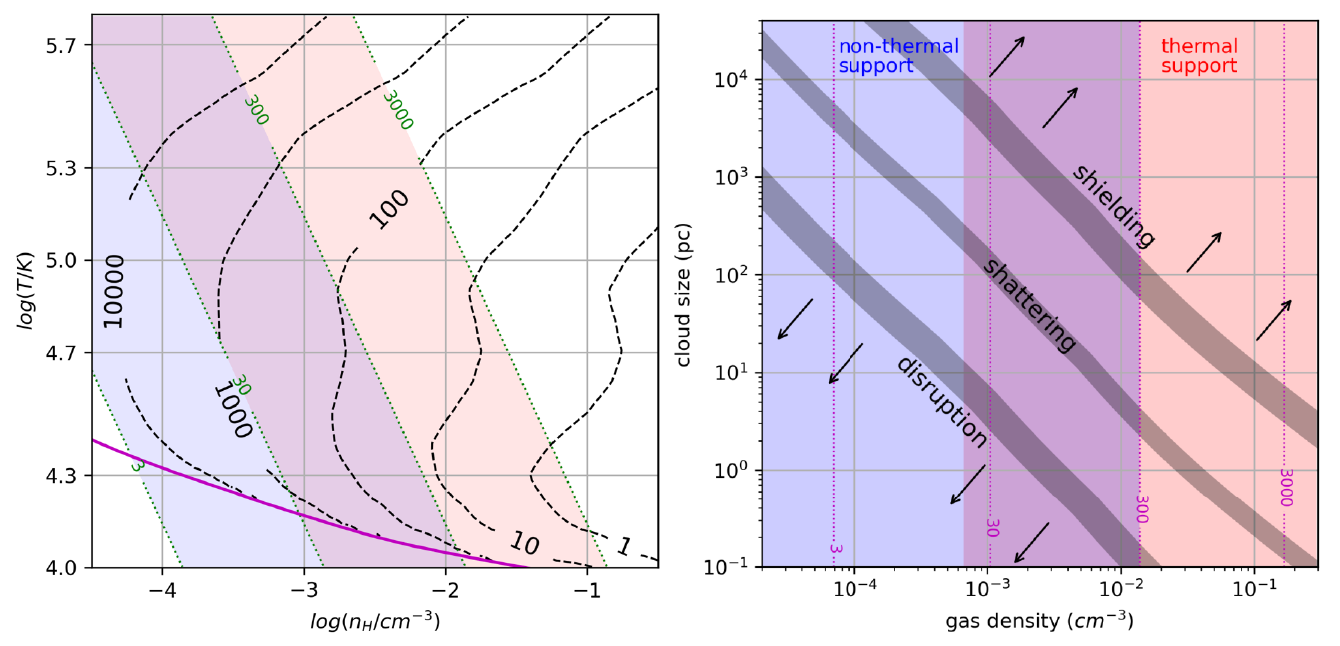}
\end{tabular}
\end{center}
\caption 
{ \label{fig:cloudScales}
\textit{Left:} Temperature-density phase space of gas undergoing thermal instability and cooling, depicting the cooling length $L_{\rm cool}$ (units of pc) in dashed contours.  Green dotted lines represent isobars (units of K cm$^{-3}$), ranges of which are shaded according to thermal (red) and non-thermal (blue) pressure support.  \textit{Right:} The cloud size vs cool gas density parameter space, showing three possible regimes : shattering, disruption, and shielding.  Shaded regions correspond to those pressures in the left panel evaluated at the photoionization equilibrium temperature for these densities.  The cloud sizes at densities relevant to the CGM ($n < 10^{-2}$ cm$^{-3}$) set our target minimum spatial resolution (10 pc) to enable imaging cloudlets in the shattering regime.}

\end{figure*}

Figure \ref{fig:cloudScales} (left) shows the cooling length in contours (units of pc) in temperature-density phase space.  Diagonal dotted lines represent isobars, and the shaded ranges of these correspond to scenarios where the cooling region is thermally or non-thermally pressure supported (red and blue, respectively) for expected CGM pressures.  Figure \ref{fig:cloudScales} (right) shows the cloud size as a function of cloud gas density where three regimes are highlighted according to the instability mechanism: shattering, the baseline size scale described above; disruption, where even smaller structures can form due to dynamical processes (turbulence and hydrodynamical instabilities, for example); and shielding, wherein some mechanism such as magnetic fields  and cosmic rays may shield and support the cooling gas, preventing fragmentation.  The implication of such small characteristic scales of cool gas in the CGM is a volume filling fraction $f_{\rm V,cool} << 1$, as the gas particles are concentrated in small clumps rather than being distributed more homogeneously throughout the medium at lower density.  \citet{FaermanWerk:2023_coolCGM} produced an analytical model of the CGM that includes non-thermal pressure support and found that $f_{\rm V,cool} \sim 1$\% is required to reproduce the observed column densities of H I and low-ionization metal species from absorption experiments.  Measuring the sizes of these structures will yield crucial constraints on the physics driving the cooling of the CGM, which is in turn needed for the fueling of star formation in galaxies. Better understanding the fundamental structure of circumgalactic gas reservoirs and the physical processes that shape them drives the need for high spatial resolution emission mapping with HWO.

A number of studies have examined the kinematic distribution of cloudlets in the CGM, and these generally agree that lines of sight in absorption studies intercept multiple clouds, often at great distances from one another, that blend together in velocity space \cite{Hummels:2017aa,Peeples:2019aa,Marra:2024_absorption}.   \citet{RubinHummels:2024_cloudflex} recently developed a parametric model to predict observational characteristics of small-scale cloud structures along lines of sight.  They find that as cloudlet sizes approach 1 pc, even velocity resolutions of 3 km/s confuse physically distinct clouds within the same velocity component.  This spectral resolution (R$\sim$100,000) is likely well beyond the reach of a high-spatial resolution IFU due to the multiplexing and optical challenges.  We note this caveat of superposed cool clouds in velocity space along any line of sight and emphasize that this both 1) further increases the importance of achieving high spatial resolution to detect variations in surface brightness over two dimensions and 2) necessitates dedicated modeling efforts to simulate the trades between spectral and spatial resolutions in discerning small-scale cool gas structure.

\section{Physical Parameters}
\label{sect:physParams} 
\subsection{Morphology of the CGM and extent of galactic superwinds}
We use existing optical emission line maps from ground-based observations, such as those described above, to exemplify the important size and velocity scales over which we should map with our UV emission line observations.  The Makani object above \citep{Rupke:2019_makani} extends over 100 kpc and exhibits two clear components, separated in velocity by 700 km/s.  The \citet{Perotta:2024_outflowingNebulae} sample exhibits emission over 40-80 kpc scales with anisotropies on scales of $<10$ kpc.  \citet{Nielsen:2024_cgmEmission} also report emission out to 30 kpc from a $z\sim0.02$ starburst galaxy, with a declining surface brightness profile that becomes substantially shallower beyond 10 kpc.  Interestingly, the [O III]/[O II] line ratios have a sharp upturn over galactocentric radii $r = 10$ to 30 kpc, suggesting the CGM rapidly becomes more ionized at these radii; mapping higher-ionization lines such as O VI at these scales is paramount to obtaining a multiphase picture of the CGM gas.

\subsection{Pressure support and CGM kinematics}
Simulations offer insight as to the velocity structure and energetics of the CGM gas.  Using the FOGGIE simulations, which feature refined resolution in the CGM, \citet{Lochhaas:2023_foggieNonHSE} find that the CGM in their models is supported by a combination of thermal pressure, rotation, turbulence, and ram pressure from bulk gas flows, with rotation and turbulence exceeding thermal pressure within 50 kpc.  The contributions from various forces induce small-scale velocity gradients over 5-10 kpc scales throughout the CGM with variations of 100 km/s over these scales.  These findings contradict common assumptions of the CGM as a purely thermal hydrostatic atmosphere surrounding galaxies but agree with results from other simulations, which also show significant support from large-scale rotation and flows \citep{Oppenheimer2018:rotatingHotHalos}. \\

\noindent\textit{Recommendations:} Large-scale UV spectral line maps should extend at least to galactocentric radii of $50$ kpc (the extent of observed and predicted superwinds, e.g., Figure \ref{fig:makani}) with $\sim$kpc scale resolution (large gas complexes in wind substructure and providing multiple resolution elements over the extent of high-resolution IFU maps).  The spaxels should enable line centroiding to $<20$ km/s (providing a dynamic range within expected velocity substructure, e.g., \cite{Lochhaas:2023_foggieNonHSE}) and cover lines from the following species (bluest to reddest): C III, Ly$\beta$, O VI, Si III, Ly$\alpha$, Si II, C IV.  These ions trace a variety of physical conditions and their line profiles and surface brightness ratios would enable inferring the gas temperature, density, and metallicity through ionization modeling.  As demonstrated in Section \ref{sect:descObs}, a UV MOS instrument with an 8' x 8' field of view (FOV) and 0.5" spatial resolution could achieve the appropriate spatial scales for systems at $z\sim0.01$.

\subsection{Small-scale structure of the CGM}
The ubiquity of absorption lines from low-ionization species tracing cool gas in the CGM suggests the presence of either large gas complexes with high volume filling factors or many smaller cloudlets with higher densities and lower volume filling factors.  The latter scenario, formed through thermal instabilities, likely encapsulates several physical mechanisms at play in the CGM, including cosmic ray support, and may be key to fueling galaxy star formation if these cloudlets precipitate and accrete onto the disk.  Alternatively, the galaxy may quench even given the presence of these clouds if they become suspended in the CGM without reaching the galaxy \cite{Thom:2012lr, Chen:2018aa, Berg:2019_LRGs}.  \\ 

\noindent\textit{Recommendations:} Figure \ref{fig:cloudScales} shows the characteristic scale of condensed cloudlets as a function of CGM physical conditions under scenarios that modulate this scale from the nominal shattering scenario.  Given characteristic cool CGM densities of $n \sim 10^{-3}$ cm$^{-3}$, high-spatial resolution spectral imaging of scales 10s-100s pc would match the expected cloud sizes (see 'Shattering' band in Figure \ref{fig:cloudScales}).  We note that this could be achieved via a high-resolution (0.05"), small  FOV (10") mode differing from the MOS used to map the large-scale winds described above.  

\onecolumn
\renewcommand{\arraystretch}{1.5}
\begin{longtable}{|p{2.9cm}|p{2.9cm}|p{2.9cm}|p{2.9cm}|p{2.9cm}|}
\caption{Primary physical parameters to be inferred in this study and the observables needed for progress at three different levels beyond the state of the art.  } \label{tab:physPars} \\
\hline
\textbf{Physical Parameter} & \textbf{State of the Art} & \textbf{Incremental Progress (Enhancing)} & \textbf{Substantial Progress (Enabling)} & \textbf{Major Progress (Breakthrough)} \\
\hline
\endfirsthead

\hline
\textbf{Physical Parameter} & \textbf{State of the Art} & \textbf{Incremental Progress (Enhancing)} & \textbf{Substantial Progress (Enabling)} & \textbf{Major Progress (Breakthrough)} \\
\hline
\endhead

1. Morphology of the CGM at low redshift& Ground-based emission mapping in H$\alpha$, H$\beta$, [O II], and [O III] of gas out to $r~\sim50$ kpc to SB of few $\times~10^{-19}$ erg/s/cm$^2$/arcsec$^2$  \citep{Nielsen:2024_cgmEmission} - densities n$_{\rm H}\sim10^{-3}$ cm$^{-3}$ & Emission mapping of hydrogen gas to equivalent SB (to State of the Art) in Ly$\alpha$ alpha emission at  multiple pointings; n$_{\rm H}\sim10^{-4}$ cm$^{-3}$ & Emission mapping of hydrogen gas to 10x fainter SB = $10^{-20}$  erg/s/cm$^2$/arcsec$^2$ in Ly$\alpha$ tiling along the minor axes; n$_{\rm H}\sim10^{-5}$ cm$^{-3}$ & Emission mapping of hydrogen gas to SB = $10^{-21}$  erg/s/cm$^2$/arcsec$^2$ in Ly$\alpha$, full map of the CGM out to the virial radius $\rightarrow$ detection of faint accretion flows; n$_{\rm H}\sim10^{-5.5}$ cm$^{-3}$
\\
\hline
2. Spatial extent of galactic superwinds & Ground-based emission maps of optical lines on scales of 100 kpc; full extent of biconical outflows 

\citep{Rupke:2019_makani} & FUV line emission maps of low-redshift galaxy halos on 30 kpc scales; disk-halo interface outflow ‘launch’ regions  & FUV line emission maps of low-redshift galaxy halos  on 60 kpc scales; half of biconical structure and galaxy disk, extent of material ejected from recent ($<$1 Gyr) starbursts

 & Ly$\alpha$, low ion (Mg II and/or Si III), and high-ion emission maps of low-redshift galaxy halos on 100 kpc scales; full extent of biconical structure and galaxy disk

\\
\hline
3. Spatially resolved phase structure of the low-redshift CGM
across ionization states  &
Ground-based emission maps of optical lines [O~II], [O~III], H$\beta$ , and H$\alpha$ 

\citep{Nielsen:2024_cgmEmission,Johnson:2022_emissionFilamentary,Zhang:2024_museCGMemission,Rupke:2019_makani,Perotta:2024_outflowingNebulae}& 
Emission maps  of low-ion $10^4$~K (Ly$\alpha$ and Si II/III) and high-ion $10^{5-6}$ K (O~VI) gas tracers & Emission maps  of multiple low- and intermediate-ion  $\sim10^4$~K (Ly$\alpha$ , Si II/III/IV, and C III) and high-ion $10^{5-6}$ K (O VI) gas tracers &
Emission maps  of FUV $\sim10^4$ K and $10^{5-6}$ K tracers  (Ly$\alpha$, Si~II/III/IV, C~III, and O~VI) gas tracers plus low-ion Mg~II emission to compare with ground-based data at low and intermediate redshift ($z\lesssim1$).  
\\
\hline
4. Sizes of small-scale structures in the CGM &
Absorption line measurements with 20 km/s velocity resolution and very rare kpc-scale separation between sightlines. 

\citep{Bowen:2016aa}& Resolve physical scales of 100 pc in CGM line emission & Resolve physical scales of few $\times10$~pc in CGM line emission & Resolve physical scales of  10 pc in CGM line emission \\
\hline
\end{longtable}
\twocolumn

\section{Description of Observations}
\label{sect:descObs} 

\subsection{Observation 1: Mapping the diffuse CGM in 12 low redshift galaxies with a range of masses and star formation activity to galactocentric radii of 50 kpc}
To ensure the highest likelihood of connecting the diffuse ionized gas in the CGM to host galaxy properties, a target sample including galaxies of varying stellar masses ($10^9 - 10^{11} \msuneq$) and SFRs is ideal.  Simulated galaxies show a range of CGM emission properties, so a sample of galaxies is required to probe a variety of populations. The target selection will also be informed by previous multiwavelength data, which enables the selection of galaxies with and without signatures of feedback, such as wind-blown extraplanar dust in edge-on galaxies \citep{HowkSavage:1997_extraplanarDustN891}. The proposed sample is four galaxies within each mass bin ($\mstareq\sim10^{9-10} \msuneq$, $\mstareq\sim10^{10-11} \msuneq$, and $\mstareq>10^{11} \msuneq$). Half of the galaxies should be star-forming and half quiescent. Of the star-forming galaxies, at least one should be nearly edge-on and at least one should be nearly face-on.  These high- and low-inclination targets will enable both clearly separated views of circumgalactic material from the disk and down-the-barrel studies \citep[e.g.,][]{Chisholm:2016aa} of galactic flows across the disk, respectively. These initial observations, while not comprising large statistical samples, will inform targeted General Observer programs to expand key regions of parameter space.

\begin{figure*}
\begin{center}
\begin{tabular}{c}
\includegraphics[width=0.75\linewidth]{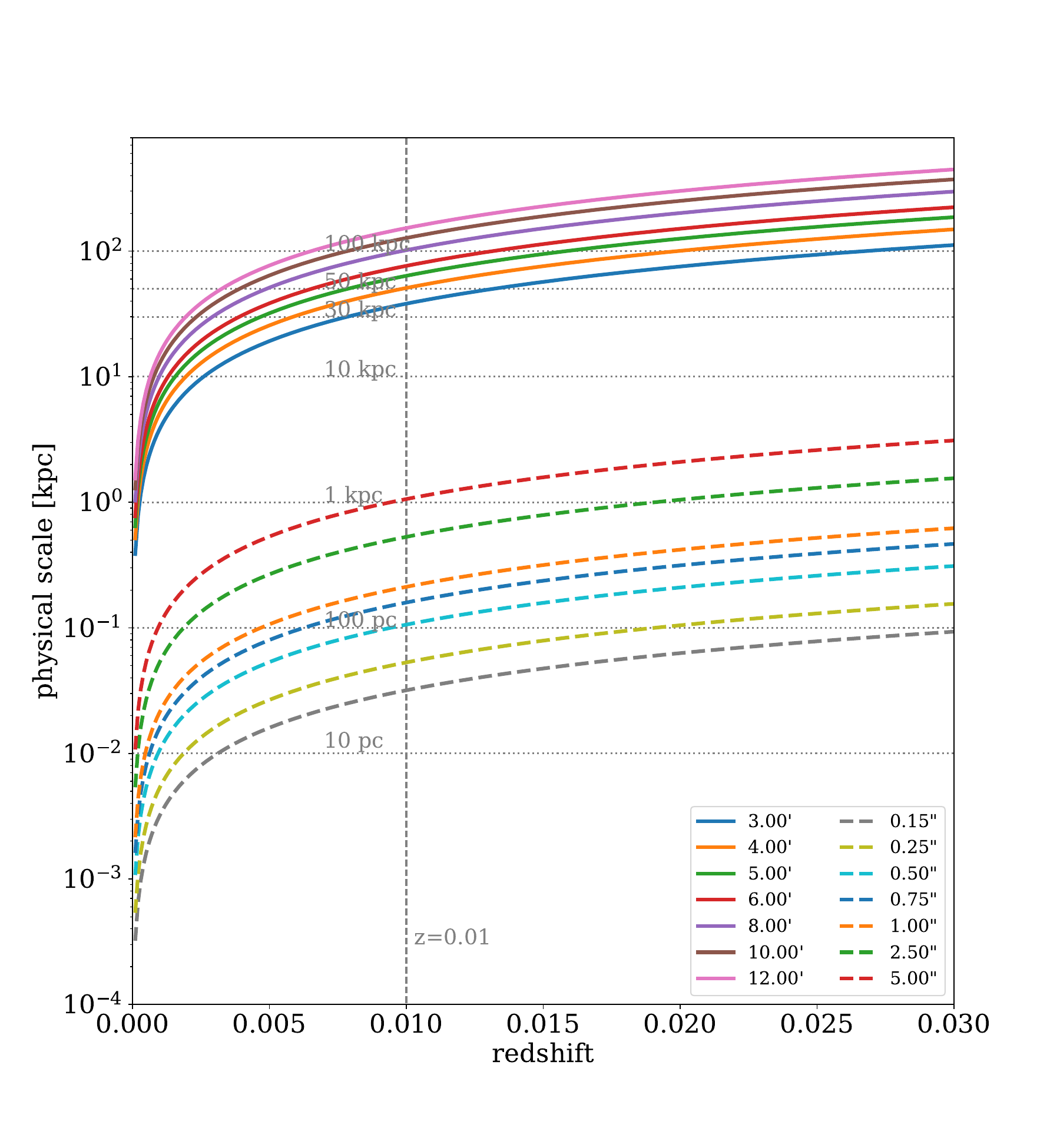}
\end{tabular}
\end{center}
\caption 
{ \label{fig:mos_fov_res}
Possible FOV and angular resolution specifications for HWO CGM spectral imaging translated to physical scales as a function of redshift.  Solid curves represent hypothetical FOV values, and dashed curves denote angular resolution values. The vertical dashed line at $z=0.01$ can guide the reader’s eye for comparisons among the instrumental characteristics for a fixed redshift akin to what should be targeted for this science case.  A FOV of 8' x 8' and resolution of 0.5" would map structures on scales of 100 kpc at 100 pc sampling.} 
\end{figure*}

We must further consider the tradeoff between FOV and spatial resolution for selecting target galaxies. Similarly, there is a tradeoff between the brightness of predicted emission from the CGM, which increases with redshift according to simulations \citep{Corlies:2016aa}, and cosmological surface brightness dimming. For this particular science case, we prioritize physical area coverage and spatial resolution over the more model-dependent concerns of redshift dependence; however, we note that programs investigating redshift dependencies will certainly be enabled by an observatory capable of delivering the science case described here.  Figure \ref{fig:mos_fov_res} shows the physical scales corresponding to FOVs and spatial resolutions engineers may consider for HWO. Solid lines show possible FOVs, and dashed lines represent spatial resolution values.  Horizontal dashed lines provide points of reference in physical scales.  We note a further technical consideration for choices of FOV and resolution: multiplexing.  The multiplex represents the total number of individual spectra produced given an FOV and angular resolution.  Ultimately, telescope/instrument design will limit the number of spectra one may capture simultaneously, and the required multiplexing for FOV/resolution configurations are provided in Table 2. 

For this observational goal, we recommend selecting galaxies at $0.003 < z \lesssim 0.01$ such that apertures would sample scales of $<1$ kpc but the FOV would approach 100 kpc, which would enable mapping key regions of the CGM with a single pointing.  

\begin{figure*}
\begin{center}
\begin{tabular}{c}
\includegraphics[width=\linewidth]{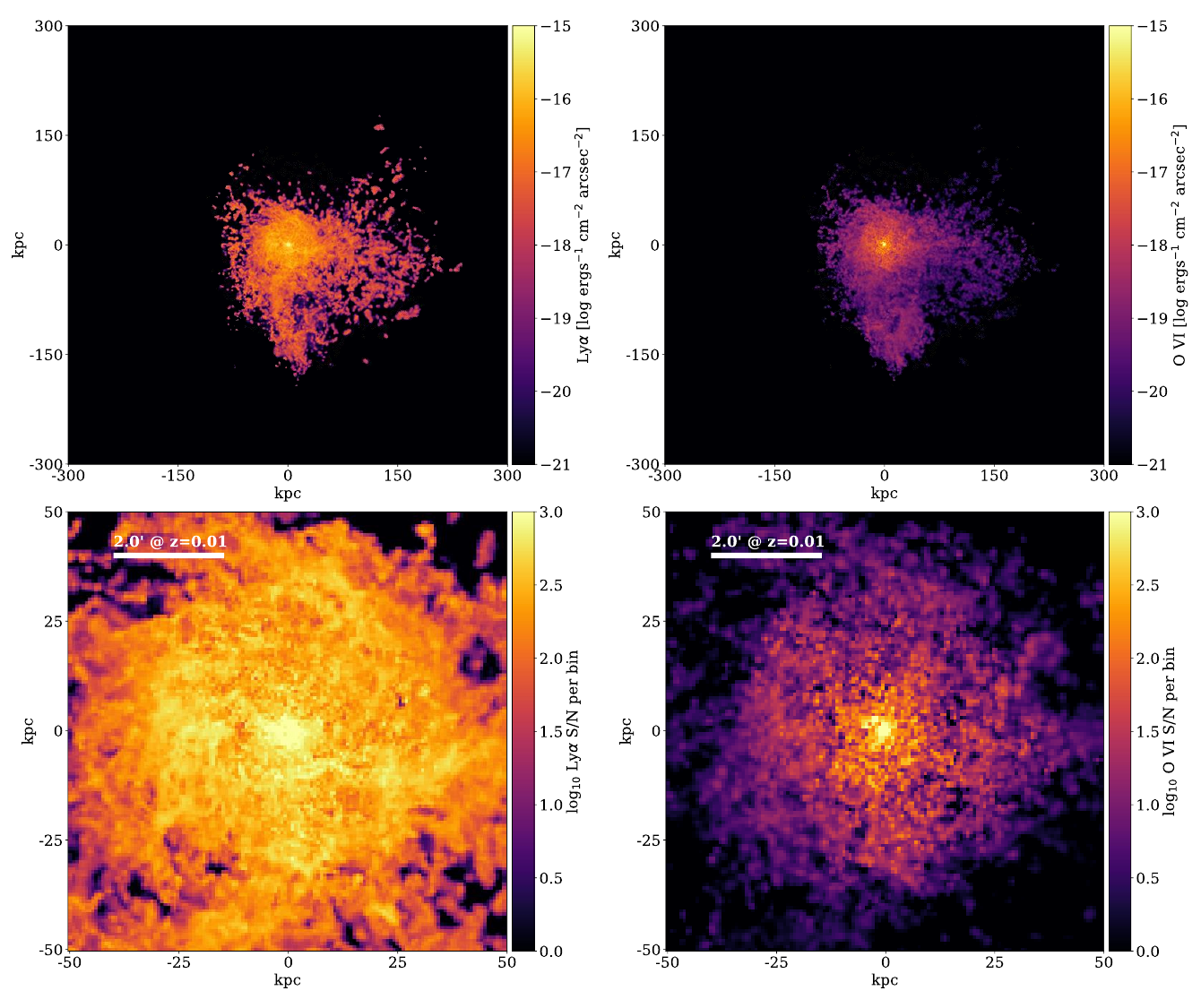}
\end{tabular}
\end{center}
\caption 
{ \label{fig:sim_sn_maps}
\textit{Top:} \lya\ and O VI emission predictions from a TNG50 simulated galaxy.  As indicated on the axes, each map is 600 x 600 kpc.  \textit{Bottom:} The signal-to-noise ratio obtained through simulated 100~ks observations of this $z=0.01$ galaxy  (redshift chosen to probe key scales; see Figures \ref{fig:mos_fov_res} and \ref{fig:ifu_fov_res}) assuming an 8-m aperture on HWO with a $8' \times 8'$ FOV UV multi-object spectrograph with binning on 0.75 kpc physical scales.  Note that the colorbars are logarithmic, indicating such an observation could reach $\sim10\sigma$ detections of CGM O VI emission out to $>30$ kpc. 
} 
\end{figure*}

For a proof-of-concept, we have utilized emission predictions from the TNG50 simulation \citep{Nelson:2019_tngDataRelease} to estimate the detectability of CGM emission.  The top panels of Figure 6 show simulation predictions of the \lya\ and O VI emission, and the bottom panels show the expected signal-to-noise ratio from a simulated 100ks observation assuming some fiducial (not official) specifications informed by current estimates of technological capabilities \citep[e.g.,][]{Tuttle:2024_hwoUvTech}.  Assumptions include an 8-m aperture with a 8' x 8' FOV micro-shutter array (MSA) UV multi-object spectrograph (MOS).  This exercise is inherently limited by the emission map from the TNG simulation, which has a spatial resolution of 750 pc. We further assume that one would open multiple shutters along the same row, which would produce overlapping spectra if an observed region had an intrinsic spectrum with many more features than the discrete emission lines assumed here. The MOS instrument used in our simulation is composed of 0.125" x 0.250" shutters, and we thus greatly oversample the pixels in the source map but bin the observation to the original 750 pc scale.  Our results indicate that $>3\sigma$ detections of O VI line emission are possible out to galactocentric radii of 50 kpc when binning to these scales.  Thus, achieving detections at higher spatial resolution is well within reach for the brighter regions.  

\subsection{Observation 2: Unveiling small-scale structure in the CGM}

While the above exercise indicates that sub-kpc scales should be within reach of HWO, there are a number of important considerations.  First, the simulated observation assumes that all shutters in the array would be kept open, which would create issues with spectra overlapping with one another, degrading the spectral resolution, and background subtraction.  These issues would be exacerbated by attempting to resolve structures with comparable angular size to the shutters themselves.  

\begin{figure*}
\begin{center}
\begin{tabular}{c}
\includegraphics[width=0.75\linewidth]{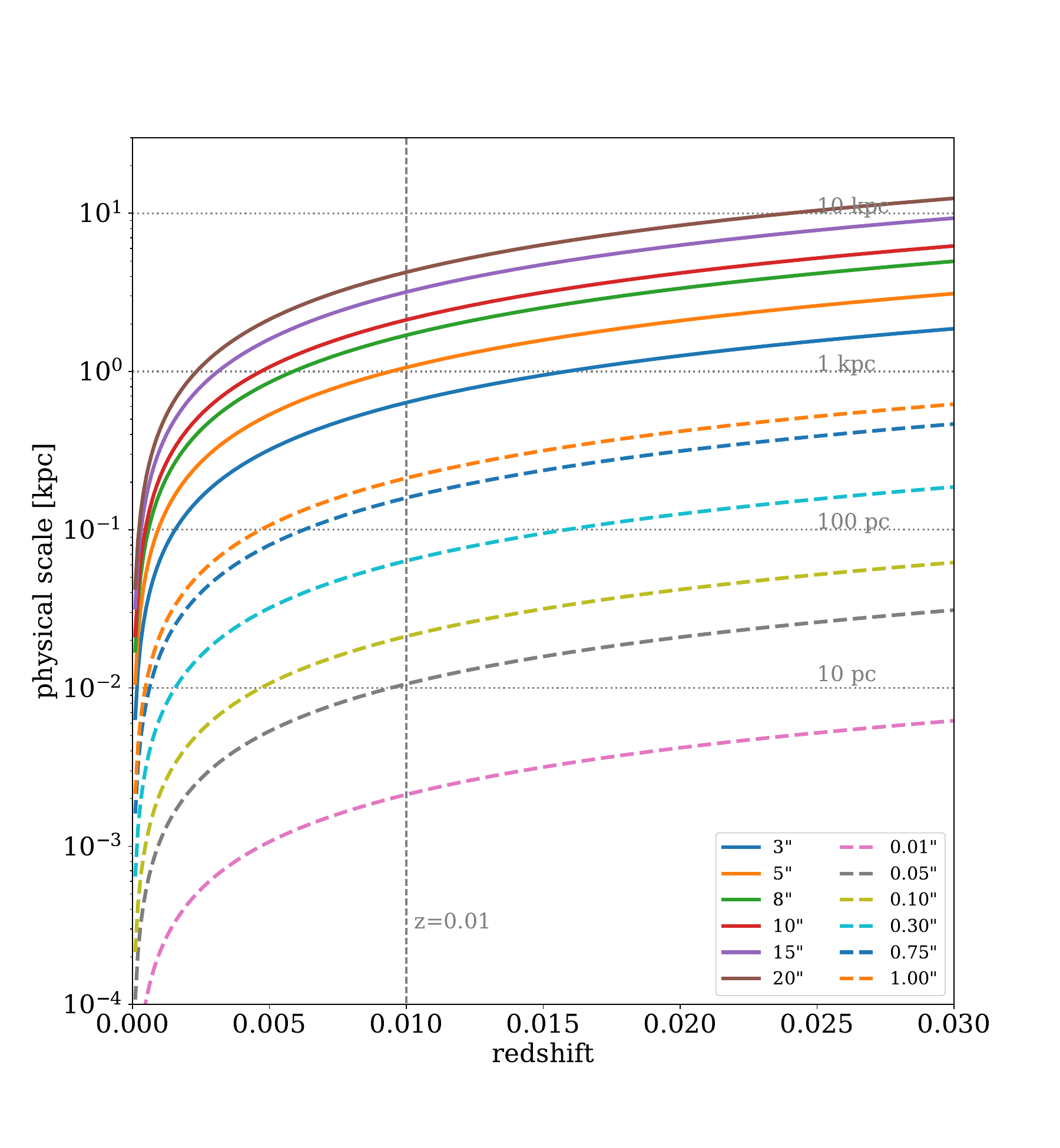}
\end{tabular}
\end{center}
\caption 
{ \label{fig:ifu_fov_res}
Similar to Figure \ref{fig:mos_fov_res} but for candidate FOV specifications for an HWO integral field unit.  Solid curves represent hypothetical FOV values, and dashed curves denote angular resolution values. If we were to follow up bright CGM emission regions discovered with the larger-FOV MOS mode at $z=0.01$, an IFU FOV of 10" would cover $\sim$3 kpc of substructure and 0.050" resolution could resolve 10 pc-scale cloudlets.
} 
\end{figure*}
Therefore, Science Objective 2 is likely better achieved with integral field spectroscopy.  An IFU designed for sub-arcsec resolution observations would enable a much greater level of control of the dispersed spectra and maximize signal for the $\sim$10s-100s pc small-scale structures we would attempt to resolve.  Observing galaxies first with the MOS mode will clearly reveal brighter regions of the CGM emission that may then be followed up with the IFU.  Figure \ref{fig:ifu_fov_res} shows a translation of candidate IFU FOVs and spatial resolutions to physical scales, analogous to Figure \ref{fig:mos_fov_res} for the MOS.  

For the same galaxy observed at $z=0.01$, which would be mapped on 100 kpc scales, CGM emission regions of interest, e.g., bright clumps, discovered with the MOS could be followed up with the IFU.  At least one kpc of substructure should be covered within the FOV of the IFU to connect with the larger scale map, and spatial resolution should reach 10 pc to resolve the smallest clouds predicted by the models above.  Thus, such an IFU should have a 10" x 10" FOV and spatial resolution of 0.050".   

The table below summarizes our observing strategy under a number of HWO mission capability scenarios.  

\onecolumn
\renewcommand{\arraystretch}{1.5}
\begin{longtable}{|p{3.0cm}|p{3.0cm}|p{3.0cm}|p{3.0cm}|p{3.0cm}|}
\caption{Specifications of HWO and the onboard UV instrument conducive to achieving the science objectives herein.  } \\
\hline
\textbf{Observation Specification} & \textbf{State of the Art} & \textbf{Incremental Progress (Enhancing)} & \textbf{Substantial Progress (Enabling)} & \textbf{Major Progress (Breakthrough)} \\
\hline
\endfirsthead

\hline
\textbf{Observation Specification} & \textbf{State of the Art} & \textbf{Incremental Progress (Enhancing)} & \textbf{Substantial Progress (Enabling)} & \textbf{Major Progress (Breakthrough)} \\
\hline
\endhead

Type (imaging, spectroscopy, etc.) & Ground-based integral field spectroscopy 

\citep{Bacon:2010aa, Henault:2003_MUSE,Morrisey:2018_KCWI,McGurk:2024_KCRM} & Micro-shutter multi-object spectroscopy (MOS) & MOS + Integral field spectroscopy (IFU) & MOS + IFU
\\
\hline
FOV & 1'~x~1'

\citep{Bacon:2010aa} &
5'~x~5'~(MOS) & 6'~x~6'~(MOS) \ 5"~x~5"~(IFU) & 8'~x~8'~(MOS) \ 10"~x~10"~(IFU)
\\
\hline
Angular resolution & 0.2" 

\citep{Bacon:2010aa} & 0.75" &0.5"~(MOS) \ 0.10"~(IFU)&0.5"~(MOS) \ 0.05"~(IFU)
\\
\hline
 Multiplex\footnotemark[1] & 90,000 
 
 \citep{Bacon:2010aa} & 400 & 720~(MOS) \ 2500~(IFU) & 960~(MOS) \ 40,000~(IFU)
\\
\hline
Spectral resolution & R$\sim$3000-13,000 

\citep{Bacon:2010aa,Morrisey:2018_KCWI} & SO 1: R$\sim$2000 SO~2: R$\sim$2000 & SO 1: R$\sim$4000 SO~2: R$\sim$4000 & SO 1: R$\sim$20,000 SO~2: R$\sim$10,000
\\
\hline
Wavelength Range & 3600-10,800 \AA\ 

\citep{Morrisey:2018_KCWI,McGurk:2024_KCRM} & 1020-1400 \AA & 1020-1400 \AA\ \& 1500-1600 \AA & 975-1600 \AA, 1500-1600 \AA, \& 2790-2900 \AA
\\
\hline
Surface brightness of target in chosen bandpass & $8 \times 10^{-19}$  erg/s/cm$^2$/arcsec$^2$ 

\citep{Nielsen:2024_cgmEmission} & $10^{-19}$ erg/s/cm$^2$/arcsec$^2$ & $10^{-20}$ erg/s/cm$^2$/arcsec$^2$ & $10^{-21}$ erg/s/cm$^2$/arcsec$^2$
\\
\hline 
\end{longtable}
\twocolumn

\section{Conclusion}
\label{sect:conc} 
Herein, we have motivated the need for UV spectral imaging capabilities on the next UV-sensitive NASA flagship mission, the Habitable Worlds Observatory, to address a key science priority for galaxy astrophysics: mapping the circumgalactic medium.  Both observations and theory point to the importance of obtaining spatially-resolved spectroscopy on both large (10s of kpc) and small (few pc) scales with the sensitivity to spectral line tracers spanning the multiple phases present from $10^4$ K to $10^6$ K.  Using current conceptual instrumental capabilities, we have demonstrated that these essential science objectives towards the larger goal of understanding Cosmic Ecosystems are well within reach of HWO given specifications outlined above.  The Great Observatory Maturation Program \citep{decadalSurvey:2021} is now entering a critical phase, where technology and engineering teams as well as project leadership will be evaluating the trade space of science goals and observatory/instrument concept specifications.  We maintain that CGM science, such as that described herein, is vital to HWO's portfolio as a general astrophysics mission.

%\acknowledgements
{\bf Acknowledgements.} J.N. Burchett is grateful for funding support from the National Science Foundation under Grant Number 2327438. We would like to thank Jessica Werk, Evan Schneider, Nicolas Tejos, and the attendees and organizers of the CGM-Chile 2024 meeting, who engaged in many invigorating conversations about this science case.  In addition, we thank Evan Scannapieco, Paul Scowen, Stephan McCandliss, and Drew Miles for critical input. 

\bibliography{Astronomy_minimal}

\end{document}